\documentclass[aps,prx,floats,twocolumn,superscriptaddress]{revtex4-2}
\usepackage{graphicx,epsfig}
\usepackage{graphics,dcolumn,bm,epic, eepic,fleqn,float}
\usepackage{amssymb,amsmath,amsfonts,multirow,rotate,color}
\usepackage{soul,xcolor,url}

\begin{document} 
\title{Metadata-informed community detection with lazy encoding using absorbing random walks}

\author{Aleix Bassolas}
\affiliation{School of Mathematical Sciences, Queen Mary University
  of London, London E1 4NS, United Kingdom}
\affiliation{Departament d'Enginyeria Informatica i Matematiques, Universitat Rovira i Virgili, 43007 Tarragona, Spain}
\author{Anton Eriksson}
\affiliation{Integrated Science Lab, Department of Physics, Ume{\aa} University, SE-901 87 Ume{\aa}, Sweden}
\author{Antoine Marot}
\affiliation{RTE Réseau de Transport d'Electricité}
\author{Martin Rosvall}
\affiliation{Integrated Science Lab, Department of Physics, Ume{\aa} University, SE-901 87 Ume{\aa}, Sweden}
\author{Vincenzo Nicosia}
\affiliation{School of Mathematical Sciences, Queen Mary University
  of London, London E1 4NS, United Kingdom}

\date{\today}

\begin{abstract}
  Integrating structural information and metadata, such as gender,
  social status, or interests, enriches networks and enables a better
  understanding of the large-scale structure of complex
  systems. However, existing approaches to metadata integration
  only consider immediately adjacent nodes, thus failing to identify
  and exploit long-range correlations between metadata and network
  structure, typical of many spatial and social
  systems. Here we show how a flow-based community-detection approach can integrate
  network information and distant metadata, providing a more nuanced
  picture of network structure and correlations. We analyse social and
  spatial networks using the map equation framework and find that
  our methodology can detect a variety of useful metadata-informed
  partitions in diverse real-world systems. This framework paves the way
  for systematically incorporating metadata in network analysis.
\end{abstract}

\maketitle

Network theory assumes that the network structure of a complex system
provides meaningful insights about its function, dynamics, and
evolution~\cite{Newman2003rev,Barrat2008,ComplexNeytworks2017}. For
example, partitioning networks into significant groups of nodes helps
researchers understanding how systems organise at different
scales~\cite{pons2005computing,newman2006modularity,lancichinetti2009community}.
But focusing only on the network topology disregards potentially
available metadata, link types or node labels that can enrich the
plain network and provide valuable information about its large-scale
organisation~\cite{peel2018multiscale,cinelli2020network}.

Researchers have used such metadata to predict missing links in
real-world networks~\cite{bliss2014evolutionary,hric2016network} and
to better characterise dynamics and
polarisation~\cite{bassolas2021first}.  Encoding link-related metadata
with multilayer networks has also proven effective for understanding
various processes in systems with diverse
relationships~\cite{DeDomenico2013,Boccaletti2014a,Battiston2017}.
A promising research direction is to integrate metadata in
community detection, the art of finding significant mesoscale
structures in networks that can guide further research to understand
the functioning of a system.

Different techniques to include exogenous information in network
clustering have been
explored~\cite{Peel2017,kitromilidis2018community}, with the aim of
generalising community-detection methods and taking account of
node-related
metadata~\cite{newman2016structure,emmons2019map,smith2016partitioning}.
For instance, extended stochastic block models and flow-based methods
can overcome the detectability limit when strong correlations between
network structure and metadata are present.  However, metadata provide
no added value in the extended stochastic block models without clear
pairwise
correlations~\cite{newman2016structure,kitromilidis2018community}.
And encoding non-aligned metadata in flow-based modules using the
extended version of the map equation further divides the structural
communities multiplicatively.  While these methods have valid use
cases, they cannot additively combine network structure and metadata,
or exploit long-range interactions to highlight either the network
structure or the metadata or any blending of the two.

We use random walks that remember their origins to enrich mesoscale
network clustering with long-range interactions between structural
information and node metadata. Assuming that each node of a graph $G$
is associated with some categorical, scalar, or vectorial metadata, we
characterise the intertwined roles of those metadata with the
structural constraints imposed by the underlying graph $G$ and detect
functional communities.  Building on the standard map equation, which
casts community detection into a compression problem, we derive a lazy
encoding scheme: we let the probability of encoding the transition of
a walker between two nodes depend on the metadata of the previously
encoded node and of the currently visited node. Equivalently, we apply
the standard map equation on the flow graph obtained by considering a
partially-absorbing random walk on $G$ that represents the probability
of coding a step on a specific node $i$. In our model, the probability
of a walker starting at node $i$ to finishing at node $j$ depends on
the node metadata at both $i$ and $j$, integrating possible long-range
interactions between structural information and metadata.  By changing
the baseline absorption probability of the walk, the proposed
framework allows us to continuously tune the relative importance of
network structure and metadata, making it easy to incorporate
field-specific knowledge in the analysis.

We show that modular compression of absorbing random walks on various
real-world networks reveals a variety of functional metadata-informed
communities. In particular, we find that many social and spatial
systems allow metadata-enriched partitions that differ substantially
from those obtained from either structural network information or
metadata clustering alone. For instance, analysing the spatial
network of energy prices across Europe reveals 
regions that do not map directly to countries or price ranges but
correspond to transnational areas characterised by
socio-economic similarities.

\section{Model}

We consider a connected and possibly weighted graph $G = (V,E)$ with
$N=|V|$ nodes and $K=|E|$ edges. For simplicity, we assume that the
graph $G$ is undirected, but a similar reasoning holds for primitive
directed graphs as well. Assuming that nodes are associated with some
categorical, scalar, or vectorial metadata such as gender, occupation,
or income, there exists a function $f:V \longrightarrow \mathcal{S}$
that maps each node $i$ to an element $f_i$ of the generic set
$\mathcal{S}$, where $\mathcal{S}\subseteq \mathbb{N}$ for categorical
data, and $\mathcal{S}\subseteq \mathbb{R}^{d}$ for scalar or
vectorial data. As shown recently, the symbolic dynamics $F(W) =
\{f_{i_0}, f_{i_i}, \ldots, f_{i_t}, \ldots\}$ associated to the
generic trajectory $W = \{i_0, i_1, \ldots, i_t, \ldots\}$ of an
unbiased random walk on $G$ starting from node $i_0$ retains plentiful
information about correlations and heterogeneity in the underlying
distribution of metadata at different
scales~\cite{Nicosia2014EPL,Bassolas2021}.

In the absence of metadata, the symbolic dynamics $F(W)$ are trivial.
Yet, the underlying random-walk statistics, including node-to-node
hitting times or the entropy rate, depend on the structure of $G$: its
degree distribution, degree-degree correlations, presence of
clustering, communities, and so forth. Several algorithms for
community detection exploit this connection between structure and
dynamics. For example, Infomap~\cite{Infomap} detects communities by
capitalising on random walks' propensity to remain trapped for
relatively long times in densely-connected subgraphs.

Conversely, if the graph $G$ presents no structural heterogeneity,
such as in an infinite lattice or a regular random graph, the
statistics of $F(W)$ depend only on the presence of correlations in
node metadata, since spurious effects due to local trapping are
averaged out in walks of infinite length.

We can interpolate between these two extremes by
considering a partially absorbing random walk whose one-step
transition probability from node~$i$ to node~$j$
is~\cite{masuda2017random}
\begin{equation}
  \pi_{ji} = \frac{w_{ij}}{\sum_{j} w_{ij}},
\end{equation}
where the link weight $w_{ij}$ between $i$ and $j$ represents the
strength of the interaction between the two nodes. Additionally, a
walker starting at node $i$ has a probability $x_{ij}$ of being
absorbed at the current node $j$, where $x_{ij}$ is some meaningful
function of the metadata $f_i$ and $f_j$. If a walker is absorbed at
$j$, its current trajectory ends at $j$. By tuning the absorption
probabilities $\{x_{ij}\}$ according to the metadata at the start node
and current node, and optimising the map equation on the resulting
absorption graph, we can include a variable amount of metadata in the
definition of meaningful communities.

The nodes' absorption probabilities represent their traversal
resistance to walkers depending on their origin and implicitly define
the walkers' horizon. For example, if $x_{ij}=1$, any walker starting
at node $i$ will always terminate as soon as it reaches node $j$. In
this sense, node $j$ presents an infinite resistance to all walkers
originating at $i$. Conversely, if $x_{ij}\simeq 0$, typically
none of the walkers starting at $i$ would ever stop at $j$. By letting
$x_{ij}$ depend on the metadata $f_i$ and $f_j$, we can drive the
walkers starting at $i$ towards nodes with specific
metadata. For example, if $x_{ij}=\delta_{f_i,f_j}$, walkers from
node $i$ will only stop at nodes whose metadata values are identical
to those at $i$.  In this case, the absorption graph will only
comprise links among nodes associated with identical metadata values,
irrespective of their actual distance on $G$, and the map equation
will be driven primarily by metadata information. Conversely, if
$x_{ij}=1$ for all $i,j$, then the absorption graph is effectively the
original graph $G$, and its structure will exclusively drive the communities. 

We define the absorption graph $\widetilde{W}=\{\widetilde{w}_{ij}\}$ from the matrix $X=\{x_{ij}\}$. The weight of the directed link $\widetilde{w}_{ij}$ between $i$ and $j$ represents the probability that a random walk that started at node $i$ is
absorbed at node $j$ after an arbitrary number of steps. The link weights can be expressed in terms of the absorption probabilities,
\begin{equation}
  \widetilde{w}_{ij} = \sum_{t=1}^{\infty} x_{ij}p_{j}(t|i),
  \label{eq:weights_i}
\end{equation}
where $p_{j}(t|i)$ is the probability for an absorbing walker that started
at node $i$ at time $0$ to visit node $j$ at time $t$. Thus, $\widetilde{w}_{ij}$ is the time integral of the probability
for a walker to be absorbed at node $j$ at time $t$ when starting from
node $i$ at time $0$. The probability of finding a walker at node $j$
at time $t$ is governed by the master equation
\begin{equation}
  p_j(t|i) = \sum_{k}p_k(t-1)\pi_{jk} (1-x_{ik}),
  \label{eq:master_node}
\end{equation}
which accounts for all the possible ways in which a walker can jump to
node $j$ at time $t-1$, given that the walker was not absorbed at time
$t-1$ on any of the neighbours of $j$. With
$\widetilde{\pi}_{jk|i} = \pi_{jk}(1-x_{ik})$ for the probability to
actually jump from node $k$ to node $j$ without being absorbed at $k$,
and $\widetilde{\Pi}_i =
\{\widetilde{\pi}_{jk|i}\}$, we can rewrite Eq.~(\ref{eq:master_node})
as
\begin{equation}
  P(t|i) = \widetilde{\Pi}_{i} P(t-1|i),
  \label{eq:master_matrix}
\end{equation}
where $P(t|i)$ is the column vector of node occupation probabilities
at time $t$ when the walk started from node $i$ at time $t=0$.  Equation~(\ref{eq:master_matrix}) is formally identical to the master
equation of a walker governed by the transition matrix
$\widetilde{\Pi}$, whose solution is
\begin{equation}
  P(t|i) = \widetilde{\Pi}^{t}_{i}P(0|i),
\end{equation}
where $p_j(0|i) = \delta_{ij}$. This means that
Eq.~(\ref{eq:weights_i}) can be rewritten as
\begin{equation}
  \widetilde{W}_i = \sum_{t=1}^{\infty}
  X_{i}^{\intercal}\widetilde{\Pi}^{t}_{i}P(0|i),
  \label{eq:flow-graph}
\end{equation}
where $X_i\> \forall i \in V$ is the column vector of absorption
probabilities for walkers starting at node $i$, and
$\widetilde{W}_{i}$ is the column vector of the edge weights of
$\widetilde{W}$ originating from node $i$. If the underlying graph $G$
is connected, the matrix $\widetilde{W}$ is in general dense since
$\widetilde{w}_{ij}$ may be non-zero even if node $i$ and node $j$ are
not directly connected by an edge.
In this sense, $\widetilde{w}_{ij}$ can be interpreted as
a conductance between $i$ and $j$.

The absorption graph $\widetilde{W}$ integrates structural
and metadata information. The structural properties of
$\widetilde{W}$ depend on the structure of the
underlying graph $G$, on the distribution of metadata across the
nodes, and on the function used to determine the absorption
probabilities $\{x_{ij}\}$. Given a graph $G$ and node metadata, the absorption probabilities
$\{x_{ij}\}$ are the only free variables. Tuning the absorption probabilities contingent upon the problem and question at hand enables a continuous dependence of $\widetilde{W}$ on $G$ and node
metadata with categorical or scalar variables.

\begin{figure*}[!htbp]
  \begin{center}
    \includegraphics[width=\textwidth]{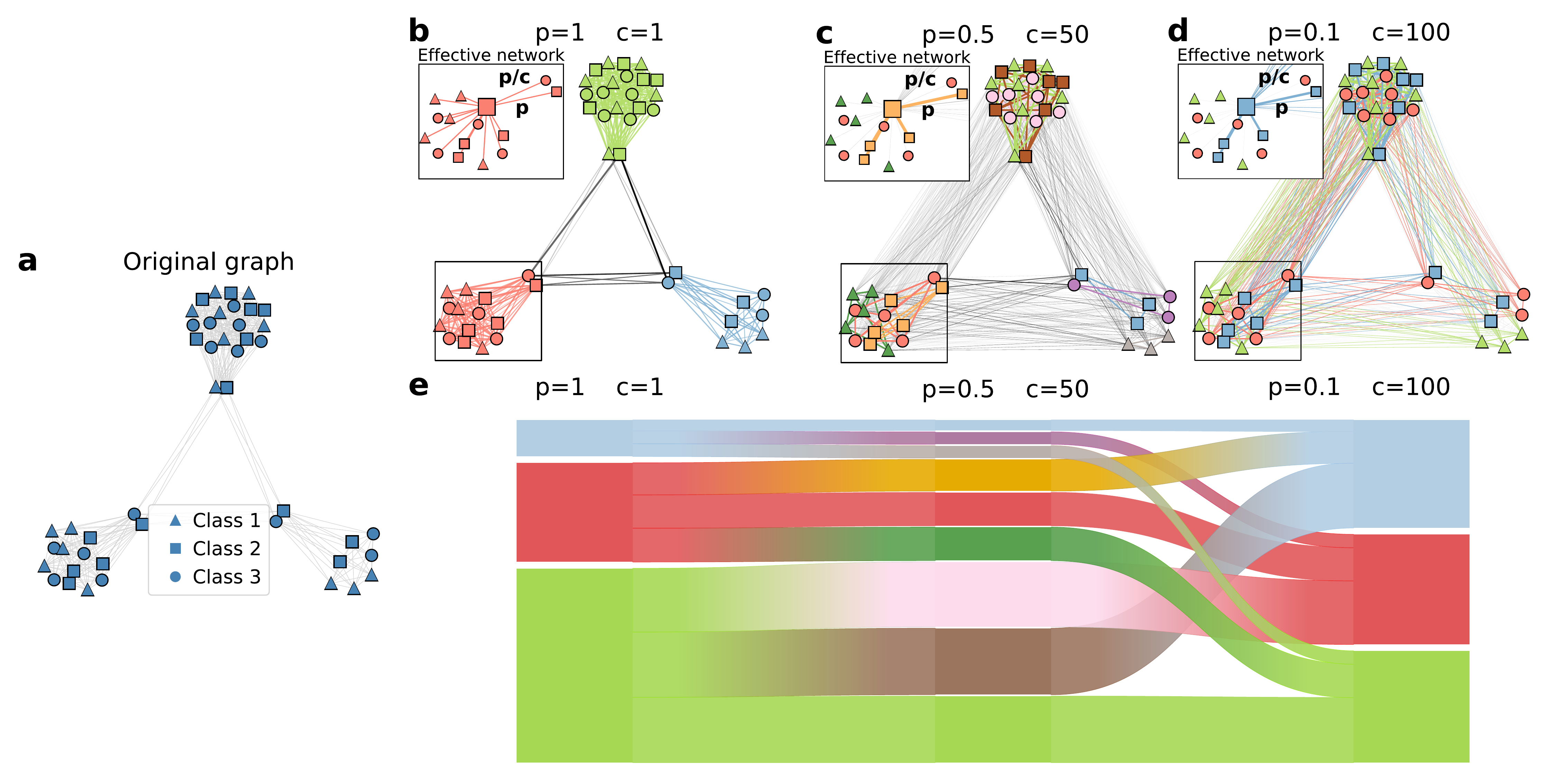}
  \end{center}
  \caption{\textbf{Metadata informed graph communities in a synthetic
      network with cliques.} \textbf{a} Original synthetic graph with
    three cliques $\mathbb{K}_{9}$, $\mathbb{K}_{15}$ and
    $\mathbb{K}_{21}$ connected by a few inter-module links. Nodes are
    associated to three categorical metadata classes, indicated by
    squares, triangles, and circles. In each clique, one-third of the
    nodes belong to each of the three classes. Each panel shows the
    links in the absorption graph corresponding to a different pair of
    values $p$ and $c$, where the inset details the probability of a
    walker to be absorbed at a node of the same class ($p$) or of a
    different class ($p/c$) of the node where it started. \textbf{b}
    Community partitions obtained in the synthetic graph for $p=1$ and
    $c=1$, which correspond to the fully connected components that
    conform the network. For such values, all transitions are encoded,
    leading to a community partition equivalent to the original
    graph. \textbf{c} Community partitions obtained for $p=0.5$ and
    $c=50$, leading to a further division of the fully
    connected cliques into three more communities determined by node
    classes. \textbf{d} Community partitions obtained for $p=0.1$ and
    $c=100$. With a lower value of $p$, nodes of the same class
    belonging to different cliques are assigned to the same community,
    yielding a partition consisting of only three communities
    determined solely by metadata. \textbf{d} Alluvial diagram depicting changes in the detected communities. In \textbf{b-d}, the links of the
    absorption graph have same colour as the nodes they connect
    if they belong to the same community, or are grey
    otherwise.} \label{fig0}
\end{figure*}

\subsection{Exploiting the map equation for absorbing random walks}

The partially absorbing random walk dynamics generalises the
traditional diffusion dynamics employed by the map
equation~\cite{rosvall2008maps}. Given a partition $\mathcal{P}$ of
the set of nodes of a graph $G$, the map equation over $G$ estimates
the average codelength needed to describe movements of the random walk
across the graph. In the original formulation, the map equation uses a
high-level index codebook to identify communities or clusters and
local modular codebooks to identify nodes inside communities. In this
way, the map equation records transitions among nodes within the same
community by using their local code words, and transitions between
nodes in two different communities with three code words: one to exit
from the current module codebook, one to enter the new module from the
index codebook, and one for the destination node from the new module
codebook.  The standard map equation records every transition and
relies only on the flow-graph defined by a uniform random walk on $G$,
and on the proposed partition $\mathcal{P}$.

When applying the map equation on the absorption graph defined in
Eq.~(\ref{eq:flow-graph}), the encoding procedure of the map equation
works as for an underlying unbiased random walk. But the
absorption graph, which represents the random walker's absorption probabilities in each node depending on its origin, will account for the
distribution of metadata in the graph. From a coding perspective, the
map equation applied to the flow-graph of absorbing random walks
corresponds to a lazy encoding scheme, where the encoding probability
depends on the metadata of the previously encoded node and the
currently visited node. Since the absorption probabilities can be tuned according to the application, this
framework represents a powerful generalisation of the classical map
equation.

\subsection{Absorption probabilities for categorical metadata}

The simplest example we consider is when each node is
associated to a binary categorical variable such as high/low,
rich/poor, $\circ$/$\bullet$, and so on, and we assume that the
probability $x_{ij}$ for a walker to be absorbed at $j$ when starting
from $i$ depends only on the categories $f_i$ and $f_j$ to which $i$
and $j$ belong, respectively. Without loss of generality, we assume
that the available categories are just $\{0,1\}$ and set:
\begin{equation}
  x_{ij} = p \delta_{f_i,f_j} +  \frac{p}{c} (1-\delta_{f_i, f_j}),
  \label{eq:absorbing_binary}
\end{equation}
where $p \in [0,1]$ and $c\in [p, +\infty]$. This assignment allows us
to model assortative, neutral, and disassortative absorption
probabilities. If $c>1$, the walker will be absorbed more frequently at
nodes belonging to the same class of the starting node (assortative
absorption), while for $p<c<1$, absorption will be more probable at
nodes belonging to a different class than the one of the starting node
(disassortative absorption). For $c = 1$, absorption does not depend
on class assignments any more (neutral absorption). Irrespective of
the value of $c$, the presence of an absorption probability $p$, which
in general is smaller than $1$, means that the walker can traverse a
large portion of the graph before being absorbed. In particular, in
the limit $p\ll 1$, the actual structure of the graph becomes less and
less relevant, and absorption is driven exclusively by categorical
information. Conversely, when $p\simeq 1$ the probability for a walker
to be absorbed at $j$ depends more on its distance from $i$ on $G$
than on metadata.  In the special case where $p=1$ and $c=1$, the
timescale of diffusion and absorption are the same, categorical
information becomes irrelevant, and the absorption graph is the
original graph $G$.

The simple synthetic graph consisting of three loosely interconnected
cliques in Fig.~\ref{fig0} shows an example of this assignment.  We
use absorption probabilities based on Eq.~(\ref{eq:absorbing_binary})
also for categorical metadata with more than two categories. We
consider real-valued metadata when the distance between two categories
can be quantified in a meaningful way and includes potentially
interesting correlations between metadata and structure.

\subsection{Absorption probabilities for real-valued metadata}

The second example we consider is real-valued node
metadata, where each node $i$ is assigned a real value $f_i$. We
assume that the absorption probability $x_{ij}$ is inversely
proportional to $|f_i - f_j|$, so that $x_{ij}$ will be higher if $i$
and $j$ are associated to similar metadata values. In general, we
could choose to modulate the absorption probability through any
decreasing function of $|f_i - f_j|$, but in the following we will
only consider the coding probability
\begin{equation}
  x_{ij} = s \exp\left(\frac{-|f_i - f_j|}{b} \right)p + (1 - s).
  \label{eq:absorption_exp}
\end{equation}
Here, $p$ is the probability to code at $j$ when $f_j=f_i$, $b\in{R}^+$
is a scale parameter, so that larger values of $b$ correspond to
higher absorption probability, and $s\in[0, 1]$ is the relative
strength of metadata information. The complement of the metadata
strength $s$ is the baseline coding probability
\begin{equation}
    s' = 1 - s,
\end{equation}
the probability to code when the distance $|f_i - f_j|$ is large.

For $s=1$, when the relative strength of metadata is
maximal, Eq.~(\ref{eq:absorption_exp}) is a proper generalisation of
Eq.~(\ref{eq:absorbing_binary}) for binary categories. In this
case, we recover $x_{ij}=p$ when $f_i = f_j$ and $x_{ij}=p/c$ when
$f_i\neq f_j$, with $c = e^{\frac{1}{b}}$.  Equivalently, for
standardised real-valued metadata with $\sigma = 1$, metadata values
that are $b$ standard deviations apart correspond to binary
separation of categories.

\begin{figure*}[!htbp]
  \begin{center}
    \includegraphics[width=\textwidth]{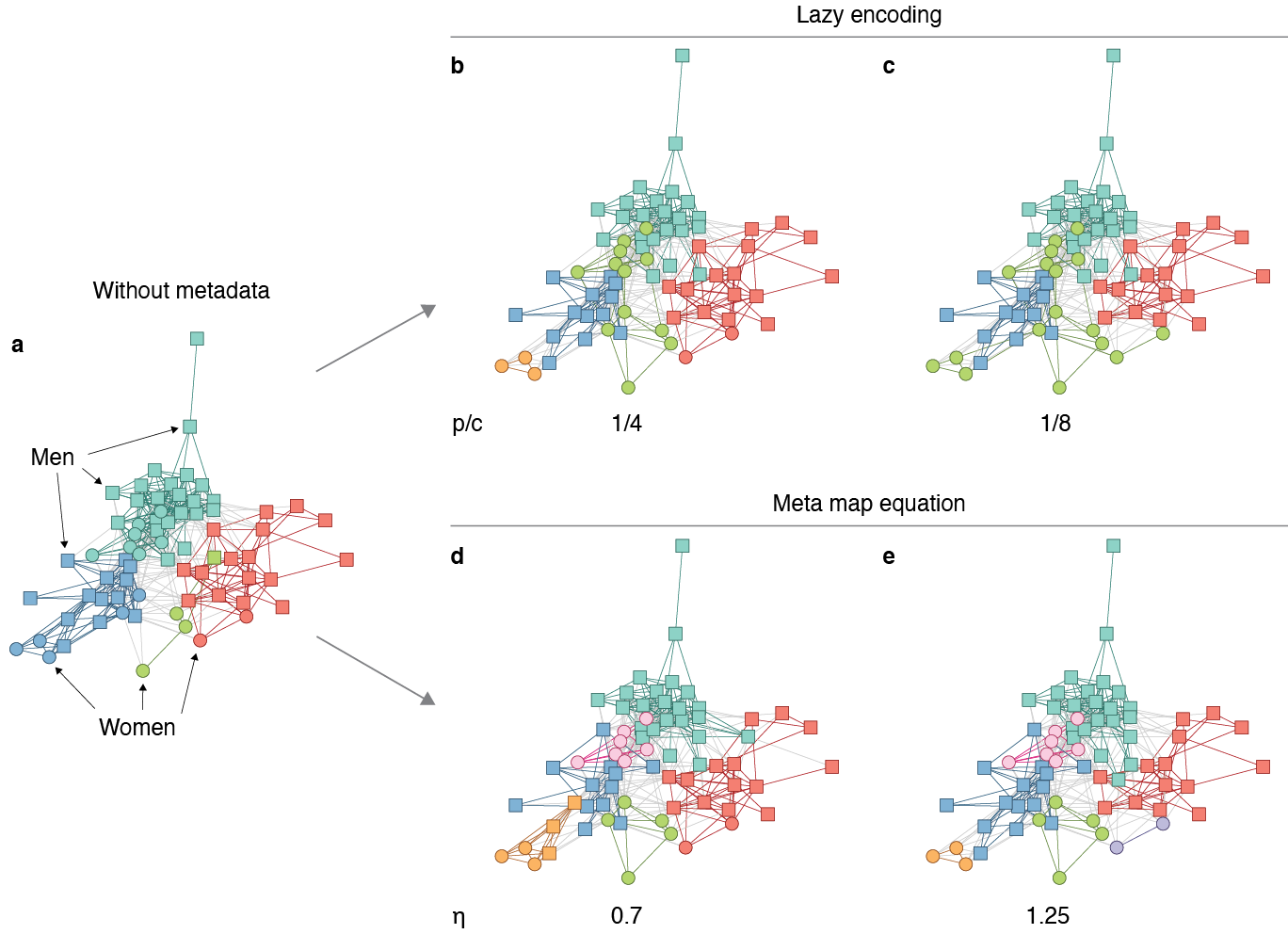}
  \end{center}
  \caption{\textbf{The optimal partitions of the Lazega lawyers'
      friendship network.} \textbf{a} With standard Infomap without gender information, the modules
    are solely determined by direct network links. \textbf{b} With
    lazy encoding random walks and encoding ratio $p/c = 1/4$, two
    women-only modules appear, and all women are included in
    three of the five modules. \textbf{c} With encoding ratio $p/c =
    1/8$, all women appear in a separate module. The map equation
    with metadata\cite{emmons2019map} gives similar modular structure for metadata
    rate $\eta = 0.8$ (\textbf{d}), which is further subdivided for metadata
    rate $\eta = 1.25$ (\textbf{e}).
    } \label{friendship}
\end{figure*}

\begin{figure}
    \centering
    \includegraphics[width=\columnwidth]{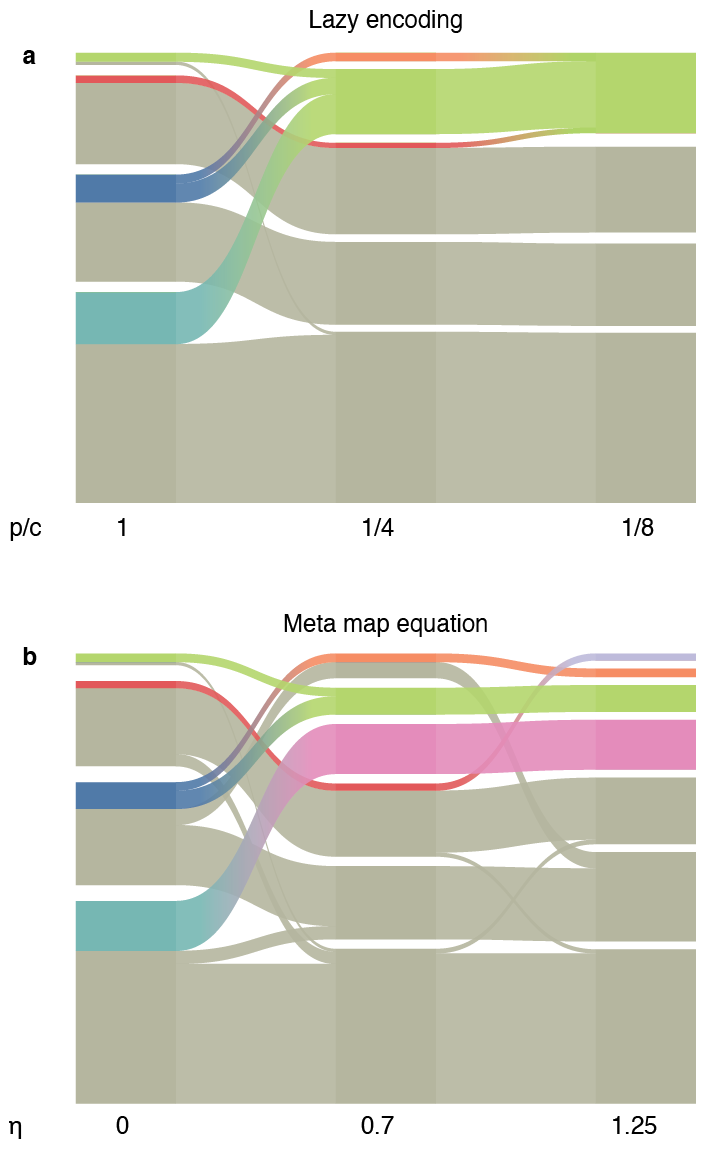}
    \caption{\textbf{Alluvial diagrams of the Lazega lawyers friendship network.} \textbf{a} With lazy encoding random walks with encoding ratios $1$, $1/4$ and $1/8$ matching panels in Fig.~\ref{friendship}\textbf{b-c}. \textbf{b} Using the meta map equation using metadata rate $\eta$ = $0$, $0.7$ and $1.25$ matching panels in Fig.~\ref{friendship}\textbf{d-e}. Encoding ratio $1$ in \textbf{a} and metadata rate $0$ in \textbf{b} discards the gender information and yields the same partition. The fraction of women in each module is coloured according to the module assignments in Fig.~\ref{friendship}.}
    \label{fig:lazegaAlluvial}
\end{figure}

\begin{figure*}[!htbp]
  \begin{center}
    \includegraphics[width=\textwidth]{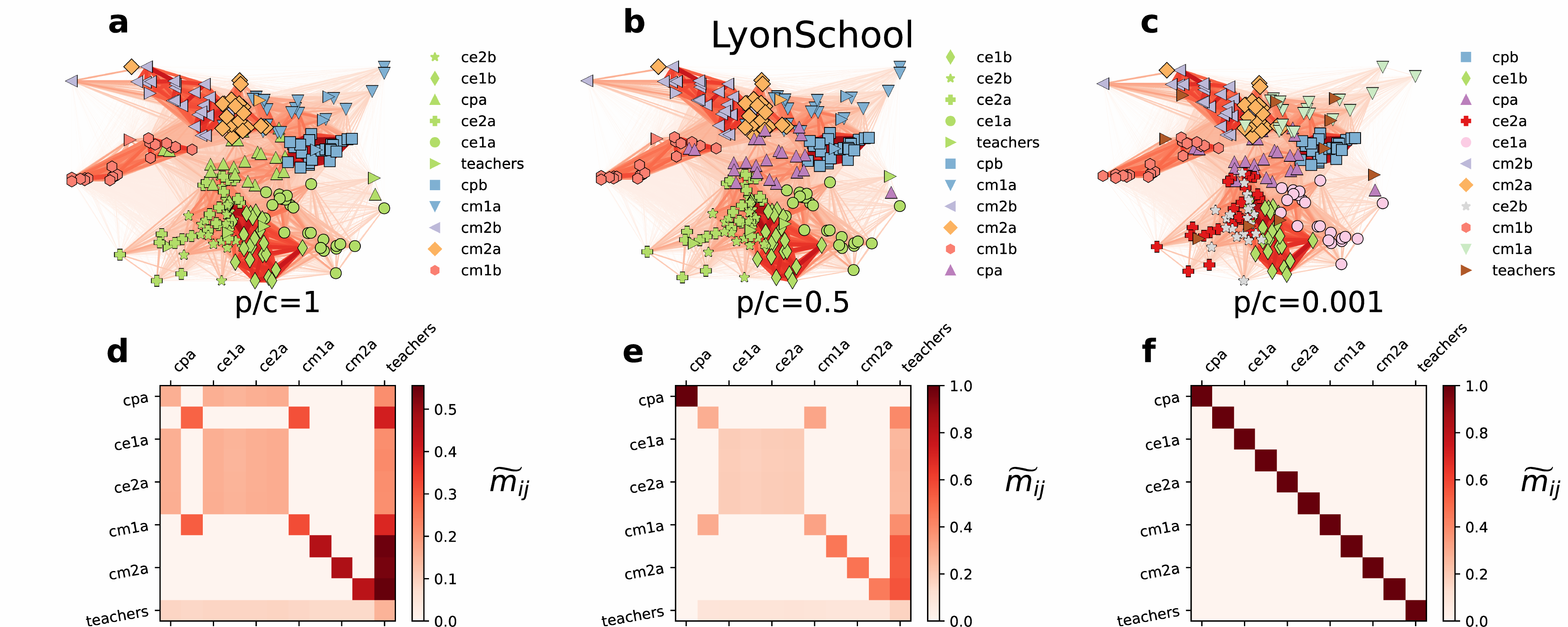}
  \end{center}
  \caption{\textbf{Metadata-based communities in the Lyon School
      contact network.} Communities in the Lyon School contact graph,
    where nodes correspond to individuals and each node is assigned a
    label corresponding to the class it belongs to. Teachers are put
    in a separate class. The probability to encode a transition is $p$
    if both nodes belong to the same class and $p/c$ otherwise. We
    show the results or $p=1$ where $c=1$ (\textbf{a}), $c=2$
    (\textbf{b}) and $c=1000$ (\textbf{c}). \textbf{d-f} Class
    overlapping assignment $\widetilde{m}_{ij}$ when $c=1$ (\textbf{d}), $c=2$
    (\textbf{e}) and $c=1000$ (\textbf{f}). Nodes are coloured
    according to their community assignment while markers indicate
    their metadata information.}\label{contact}
\end{figure*}

\section{Results}

We analysed the range of community partitions found by the map
equation's search algorithm Infomap~\cite{Infomap} on the absorption
graphs of various synthetic and real-world systems. For each network,
we constructed the absorption graph of Eq.~(\ref{eq:absorbing_binary})
for different values of $p$ and $c$.  To illustrate how the absorption
graph can integrate metadata and structural information, we start from
a simple example: an unweighted synthetic network with three different
classes equally distributed in three fully connected subgraphs with
only a few links between them (Fig.~\ref{fig0}\textbf{a}).  In
Fig.~\ref{fig0}\textbf{b}-\textbf{d}, we show the partitions in
communities identified by Infomap on the absorption graphs obtained
for different values of $p$ and $c$, corresponding to different ways
of mixing structural and metadata information. When $p=1$ and $c=1$
(panel \textbf{b}), metadata play no role, and the best partition
consists of three communities, corresponding to the three cliques. As
expected, this partition is identical to the one obtained running
Infomap on the original graph.  When $p=1$ and $c=1$ the absorption
graph is identical to the original graph $G$.  Varying $p$ and $c$
strikes a balance between metadata and structure, obtaining other
meaningful partitions. When $p=0.5$ and $c=50$, our method reveals a
total of $9$ communities instead of just $3$
(Fig. \ref{fig0}\textbf{c}).  Each clique has been split in three
sub-modules, corresponding to the nodes' metadata assignments. In
general, for large values of $p$, walkers are absorbed before having
the chance to move between communities. However, for sufficiently
small values of $p$, walkers are able to move to other cliques, and to
``see'' other nodes belonging to the same class as that of their
original node. This leads to another meaningful partition where nodes
are grouped by metadata values (Fig.~\ref{fig0}\textbf{d}).

\subsection{Social contact networks}

Several real world contact networks have metadata attached to nodes,
providing explicit information about the function or position of any
given individual in the system. For instance, the metadata can
identify the role of each node of a hospital contact network, or the
class to which students of a school belong. Taking into account this
information can be crucial to correctly interpret how a system
functions.

As a first example we considered the Lazega lawyers friendship
network~\cite{lazega2001collegial} with gender information.  We report
in Fig.~\ref{friendship} the partitions obtained by using different
values of the parameters $p$ and $c$. In panel \textbf{a}, $p=1$ and
$c=1$, so that no gender information is considered, and the partition
is purely relying on the structure of the graph. In panel \textbf{b}
and \textbf{c}, instead, the gender of each node increases in
relevancy, leading to an almost complete separation between nodes of
different genders in \textbf{c}. The transition between
structure-focused and metadata-focused partitions is illustrated in
the alluvial diagram in
Fig.~\ref{fig:lazegaAlluvial}\textbf{a}. Infomap puts female nodes in
a community of their own for sufficiently low values of $p/c$,
irrespective of the original structural cluster they belonged to.  To
compare our results with previous work, we reproduced the results
obtained using the meta map equation using metadata rate $\eta = 0.7$
and $1.25$ (ref.~\cite{emmons2019map}, Fig.~3).  In panel~\textbf{d},
the women start to separate from the men, but in different
ways. Overall, increasing the meta data rate separates men and women,
but tends not to join back clusters consisting of nodes of the same
gender. When the metadata rate $\eta = 1.25$ (panel \textbf{e}), men
and women are completely separated, but same-gendered modules are
never joined again. As before, the alluvial diagram in
Fig.~\ref{fig:lazegaAlluvial}\textbf{b} shows how small sub-clusters
of female nodes are isolated from larger clusters of male ones, and
never aggregated.

As another example of metadata-enriched social networks, we consider
the Lyon School contact graph~\cite{genois2018can}, which is one of
several annotated social networks made available by the SocioPatterns
project~\cite{sociopatterns}. The graph reports the face-to-face
interactions among students in a school of Lyon. For each node, we
know whether the corresponding person is a student or a teacher and
to which class she belongs. In total, the data set consists of a dense
network with $N=242$ nodes, $K=26594$ links, and eleven node classes
(10 classrooms plus teachers).  In Fig.~\ref{contact}, we display the
partitions obtained for $p=1$ and $c=1$ (\textbf{a}), $c=2$
(\textbf{b}), and $c=1000$ (\textbf{c}). By increasing
$c$, the class of each node becomes more relevant and the structural
communities start to split. Teachers are the last metadata class to be
recovered because the majority of face-to-face contacts of each
teacher happen with pupils in their respective class. The matrices of
class overlap $\widetilde{m}_{ij}$ (see Methods for details) clarifies
the role of $c$, and shows how all nodes assigned to the same class
end up in the same community when $c$ is sufficiently large and the
probability for a walker to be absorbed at any node is relatively
small. For additional results using two other social networks, see
Supplementary Figs.\ S-1 and S-2.

\begin{figure*}[!htbp]
  \begin{center}
    \includegraphics[width=\textwidth]{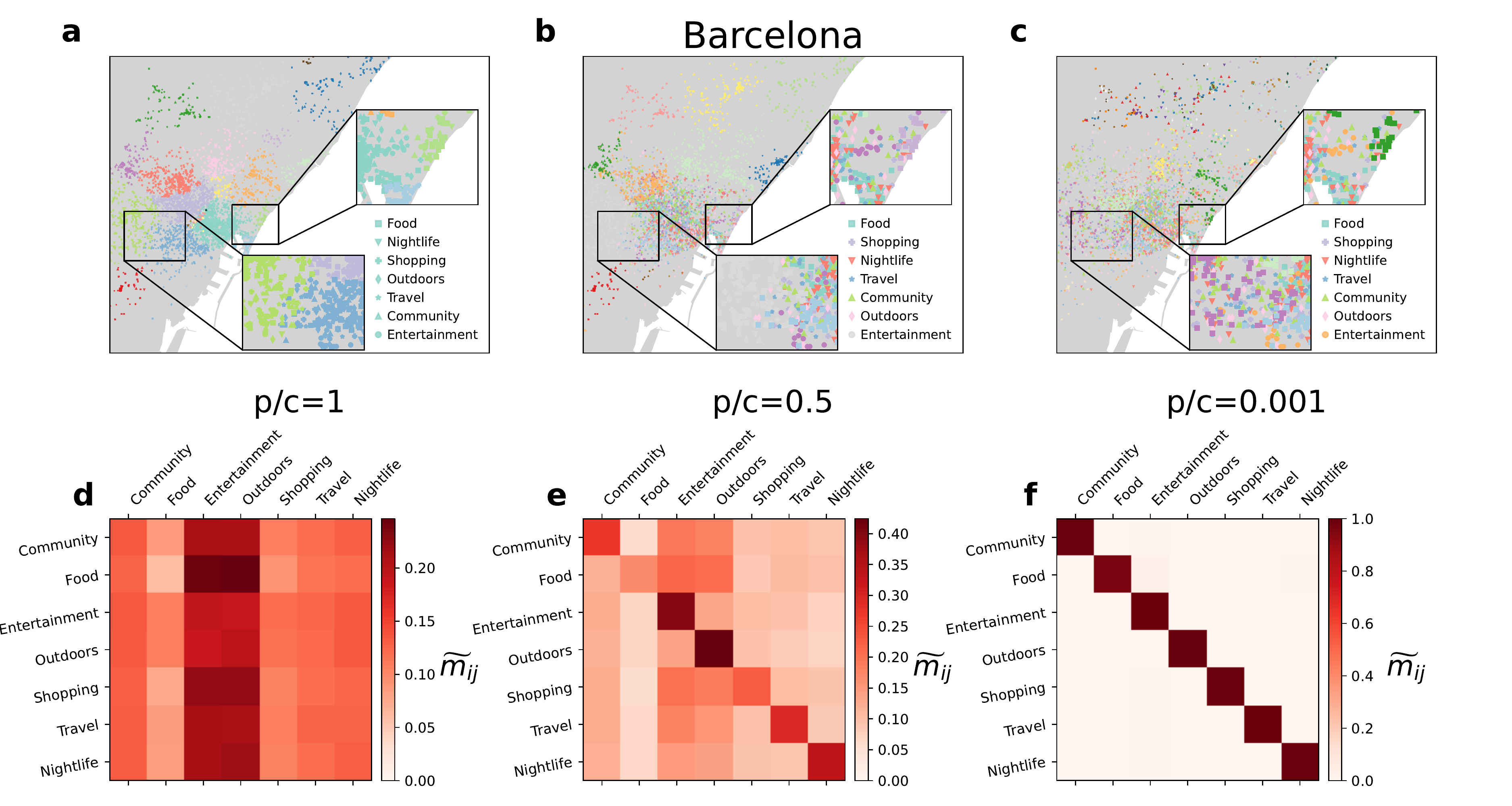}
  \end{center}
 \caption{\textbf{Communities of Gowalla venues in Barcelona.}  The
   venues in Barcelona tracked by Gowalla user activity form a spatial
   graph where any pair of venues is connected by a link if they are
   less than $2$ km apart. Nodes are divided into six classes,
   according to the type of venue. The partitions obtained for $p=0.5$
   and $c=1$ (\textbf{a}), $c=2$ (\textbf{b}) and $c=1000$
   (\textbf{c}) are reported. \textbf{d-f} Class overlapping
   $\widetilde{m}_{ij}$ when $c=1$ (\textbf{d}), $c=2$
   (\textbf{e}) and $c=1000$ (\textbf{f}). Venues are coloured
   according to their community assignment and the marker indicates
   the type of venue.}\label{gowalla}
\end{figure*}

\begin{figure*}[!htbp]
  \begin{center}
    \includegraphics[width=\textwidth]{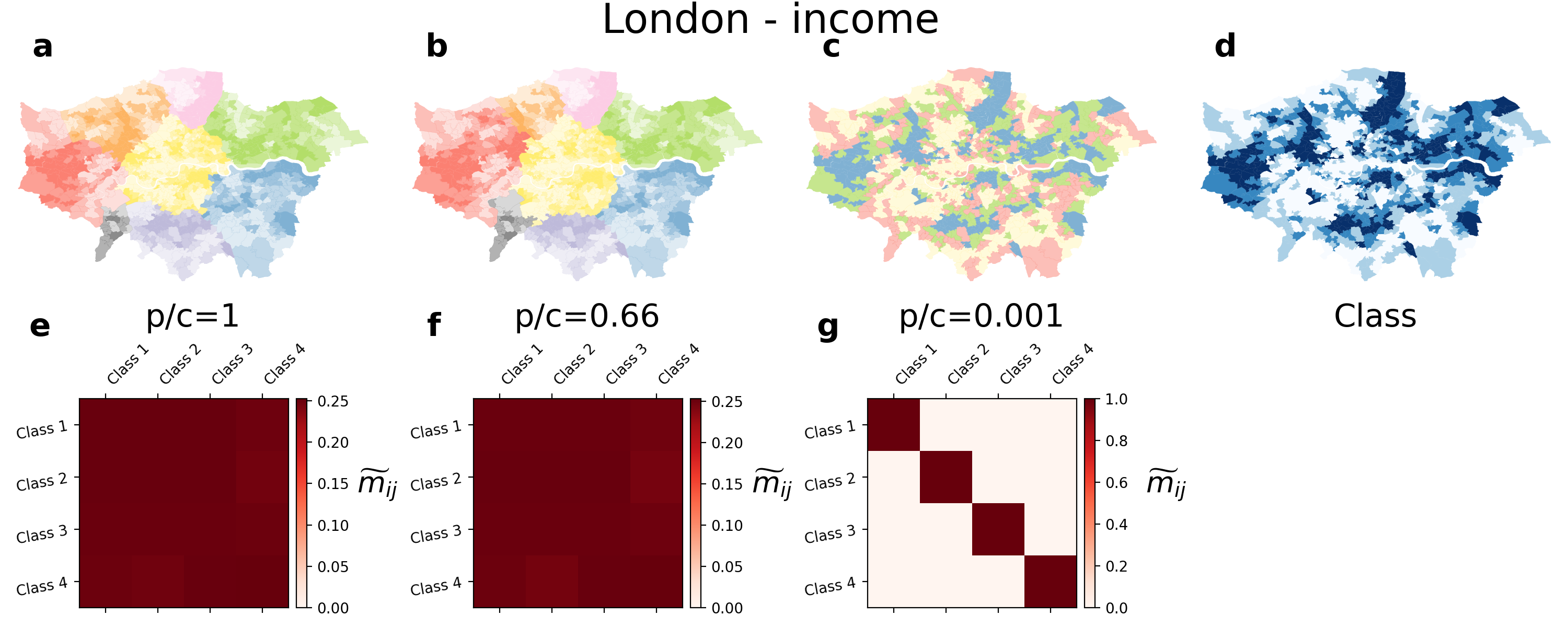}
  \end{center}
  \caption{\textbf{Partition of areas of London according to income
      quartiles.} Each node here is an MSOA region of Greater London,
    two regions are connected by a weighted link corresponding to the
    commuting flow between them, and the metadata is income quartile,
    where regions in class 1 are the poorest and regions in class 4
    are the most wealthy, respectively. The different partitions
    correspond to $p=1$, and, respectively, $c=1$ (\textbf{a}), $c=2$
    (\textbf{b}) and $c=1000$ (\textbf{c}), with regions coloured
    according to their community assignment. \textbf{d}
    The class assignment of each region.  \textbf{e-g} Class
    overlapping $\widetilde{m}_{ij}$ when $c=1$ (\textbf{e}), $c=1.5$
    (\textbf{f}) and $c=1000$ (\textbf{g}).}\label{london}
\end{figure*}

\subsection{Organisation of activities in urban areas}

The proposed methodology can help identify functional
modules in spatial systems. Standard community detection algorithms
often do not provide the desired results on spatially-embedded
networks. The spatial constraints are too strong to allow
communities whose nodes are too far apart from each other. However,
metadata is often available in spatial networks, and taking this
information into account when detecting modules is desirable in many
concrete applications. Typical examples include analysing spatial
correlation in the distribution of certain commercial activities or
identifying spatial segregation according to a specific socio-economic
indicator.

We consider a spatial data set constructed from the location-based
social network Gowalla~\cite{cho2011friendship,liu2014exploiting},
which includes the location and type of millions of venues across the
world. Whereas in this data set each venue has multiple classes
organised in a hierarchical way, we have only analysed the main six
categories: food, nightlife, outdoors, community, entertainment and
travel. The graph connecting the venues is spatial so that there is a
link between any pair of venues if the distance $d_{ij}$ separating
them is lower than $2$~km and the weight of each link is given by
$\log(1/d_{ij})$\cite{Noulas2011}. Figure~\ref{gowalla} shows the
partitions obtained on the network of commercial activities in
Barcelona for $p=0.5$ and $c=1$ (\textbf{a}), $c=2$ (\textbf{b}) and
$c=1000$ (\textbf{c}). For $c=1$, the venues are organised in spatial
communities, determined solely by the relative distance among
nodes. Already for $c=2$, some of the communities split out, leading
to a grouping of venues of the same type. Still, the more isolated
spatial communities do not split out until $c=1000$, where the vast
majority of venues of the same category are clustered
together. Whereas the results shown in Fig. \ref{gowalla} correspond
to $p=0.5$, by changing $p$ we can also tune the typical size of the
spatial communities, with higher values leading to smaller groups.
For additional results using three other cities, see Supplementary
Figs.\ S-3 and S-4.

\subsection{Income distribution and mobility in London}

In spatial systems, networks can be defined in multiple ways, not only
according to the distance separating two regions, but also according
to the number of people that move between places. We show the
partitions in communities obtained for the mobility network of London,
where the median household income of an area is the relevant metadata.
We considered Greater London at the level of Middle Layer Super Output
Areas (MSOAs), where each MSOA is a node and the weighted network
connecting MSOAs is based on the mobility of commuters (See Methods
for details).  Each node is associated to one of four classes
according to the median income of the households in the corresponding
region.  Each class represents a quartile of the median income
distribution, with class $1$ being most deprived and class $4$ the
wealthiest. The results are shown in Fig.~\ref{london} for $p=1$ and
$c=1$ (\textbf{a}), $c=2$ (\textbf{b}) and $c=1000$ (\textbf{c}). By
increasing the relative importance of metadata
(Fig.\ref{london}\textbf{d}), we obtained significant changes in the
partitions. In particular, when $c$ is small, the modules are
effectively determined mainly by spatial distance, while only four
communities based on metadata information are obtained when
$c=1000$. The changes in the partitions can be more clearly observed
in Fig.~\ref{london}\textbf{e-g}, where the increase of $c$ affects
the number of MSOAs of the same income category assigned to the same
community. In Supplementary Figs.\ S-5, S-6, S-7, and S-8 we show
similar results for other metadata quantities.

\subsection{Power grid network}

\begin{figure*}[!htbp]
    \centering
    \includegraphics[width=\textwidth]{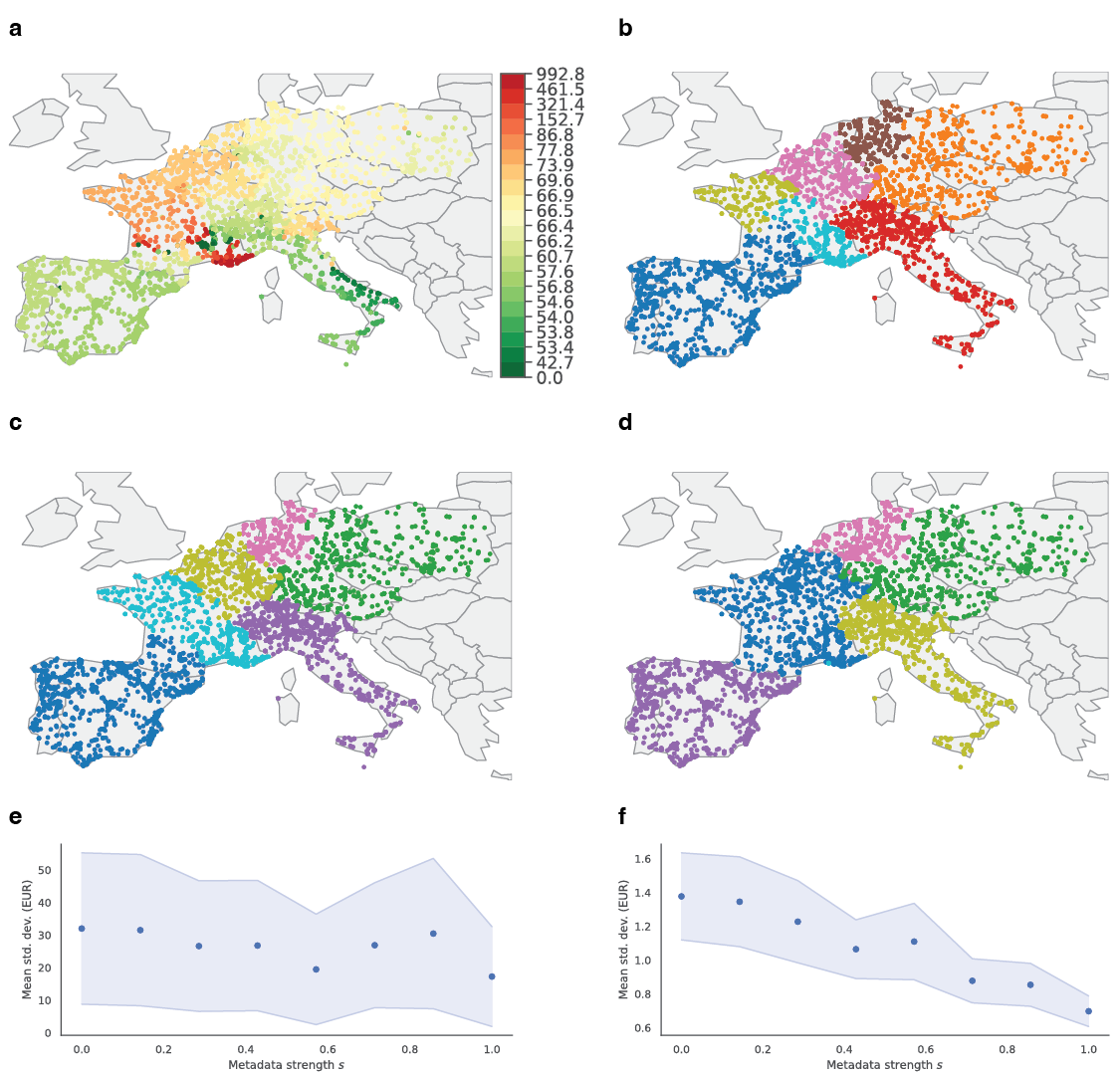}
    \caption{\textbf{European power grid network with node prices and optimal partitions.} The partitions in panels \textbf{b-d} are hierarchical with 7 levels. Here, we show only the top-level super-modules.
    \textbf{a} Node prices (EUR) distribute with lower prices in southern and central Europe and higher in western Europe.
    \textbf{b} With metadata strength $s=0$, modules contain high- and low-price regions in the partition resulting from only the network structure.
    In panels \textbf{c-d}, node prices influence the partitions. \textbf{c} With $s=0.3$, the majority of western Europe is divided into only two modules. \textbf{d} With $s=1$, the border between low-price southern Europe and higher-priced western Europe becomes visible. In panels \textbf{e-f}, the mean of module price's standard deviation for the top-level (\textbf{e}) and leaf-level (\textbf{f}) for different metadata strengths. The coloured bands represent the standard error of the mean.}\label{fig:powergrid}
\end{figure*}

As a final example, we consider the European electrical power grid,
which forms a transport network connecting electricity producers and
consumers. This system is structured in a similar way to a road
transport network, with ``highways'' between large hubs, connecting
smaller neighbouring cities. Yet electricity can flow instantly all across Europe, making it a large and dynamic system to consider as a whole. Each node has an associated electricity
price. In the analysed situation, the price distribution is relatively heterogeneous, with
low-price regions mainly around southern Europe with high solar production, average-price
ones in central Europe, and high-price regions in western Europe
(Fig.~\ref{fig:powergrid}\textbf{a}). Overall, prices are
correlated in space, but price ranges do not necessarily map into
countries or other political divisions. Bidding zones for energy markets were historically defined by country borders. However, the European Institutions wants to revise them to make energy markets more efficient over more coherent and connected  price zones, or communities, that do not depend on country borders.

To include real-valued prices as metadata in the lazy encoding random
walk, we derived the coding probability using the price distance
between nodes.  We started $10^7$ random walks per node to achieve
ergodic visit rates, repeating the simulation with metadata strength $s$
varying between $0$ and $1$.  We ran 100 optimisation trials with Infomap for each
simulation and chose the partition with the lowest codelength.

The resulting partitions have six or seven levels of nested modules,
organised into six or seven top-level super modules (one module with
four exceptionally high price nodes in Fig.~\ref{fig:powergrid}\textbf{d}).  Western Europe
organises into fewer modules with increasing metadata strength $s$, and the southern-western Europe border appear
(Fig.~\ref{fig:powergrid}\textbf{b-d}).  With increasing metadata strength,
the mean module-price variance decreases
monotonically on the leaf level (Fig.~\ref{fig:powergrid}\textbf{f}).
On the top level, the behaviour is erratic but overall
trending towards lower variance
(Fig.~\ref{fig:powergrid}\textbf{e}). Overall, incorporating
metadata in the community detection task provided a more
nuanced picture of the correlations between energy price and geography beyond country borders while still defining connected zones.

\section{Discussion}

Integrating structural and metadata information beyond nodes' first
neighbours has been a standing challenge in network-based data
analysis. We have shown how to include metadata in the map
equation with metadata-dependent absorbing random walks. By coupling
the absorption probability to the metadata of nodes, the absorbing
random walk dynamics produce a tunable lazy encoding scheme: The
metadata modulate a walker's coding horizon, simultaneously accounting
for the node classes' distribution and their correlations in the
network. This approach equips researchers with a tool for identifying
mesoscale structures in networks based on link structure and discrete
or continuous node metadata information.

Often metadata about the components of a system are as relevant as
knowing how those components connect.  A researcher must use her
application-specific knowledge when deciding how much metadata
information she should include when identifying functional modules of
a complex system. Few algorithms for community detection consider
metadata~\cite{Peel2017,emmons2019map}, and those that do impose heavy
constraints either on the relative role of structure and metadata or
on the way they use metadata. For instance, current algorithms for
detecting flow-based communities further divide communities into
smaller metadata-based sub-clusters.

To enable related sub-clusters to merge or groups of nodes to move
from one cluster to another based on their metadata, we propose a
simple formalism to modify the encoding procedure of the map
equation. Metadata-dependent absorption probabilities at each node
induce an absorption graph that encodes the relative importance of
structure and metadata. In general one
could use any community-detection algorithm for weighted and directed
links on the absorption graph, and interpret the results. However,
using the map equation's search algorithm Infomap provides a
principled interpretation: A tunable lazy encoding scheme, which
extends and generalises the standard map equation formalism in a
natural way.

Tunable absorption probabilities allow the researcher to incorporate
specific field knowledge in community detection easily. In the
examples, we have shown how different values of $p$ and $c$ give
various relevant solutions. In general, larger values of $c$ let
walkers visit larger portions of the graphs before being absorbed,
allowing for relatively distant nodes with similar metadata to be
clustered together. Similarly, smaller values of $p$ tend to yield
larger clusters. A single recipe for setting $p$ and $c$ is not only
unavailable but also not needed. A user benefits from exploring the
parameter space and selecting the most meaningful ranges of $p$ and
$c$ that provide informative partitions.

Linking network structure and metadata through the dynamics
of random walks on networks in our compression-based approach opens a
new avenue for community detection. Various types of information about
nodes and edges, such as the physical location of nodes, edge classes,
or other exogenous classification and rankings, provide enticing
directions to explore for new insights about complex systems.

\section{Methods}

\subsection{Class overlap}

We assess the extent to which nodes with different categorical metadata information
are assigned to the different (or the same) communities by proposing the class overlap $m_{\alpha \beta}$ between classes $\alpha$ and $\beta$. The class overlap
\begin{equation}
  m_{\alpha \beta} = \frac{1}{N_{\alpha}+N_{\beta}} \sum_{\forall i \in \mathcal{C}} N_{i,{\alpha}} + N_{i,{\beta}},
\end{equation}
 where $\mathcal{C}$ are the communities reported by our algorithm, $N_{\alpha}$ is the total number of nodes of class $\alpha$, and $N_{i,\alpha}$ is the total number of nodes of class $\alpha$ in module $i$. 

This quantity is equal to one when the classes $\alpha$ and $\beta$ are evenly divided and zero when the nodes of those classes are in separate communities. For easy comparison between different classes, the final quantity we have considered is its row-normalised counterpart given by

\begin{equation}
  \widetilde{m}_{\alpha \beta} = \frac{m_{\alpha \beta}}{\sum_{\beta} m_{\alpha \beta}}.
\end{equation}

This metric assesses how nodes with different (or equal) metadata information
are assigned to the same community when the lazy encoding is tuned.

\subsection{Commuting network in London}

The commuting data analysed in the income and mobility section (see data availability) includes the number of individuals $\mathcal{T}_{ij}$ that live in a spatial unit (MSOA) $i$ and work in
a spatial unit $j$. From those commuting patterns, we have built the mobility network between MSOAs in which the weight of a link going from a spatial unit $i$ to a spatial unit $j$ is given by $w_{ij}=\mathcal{T}_{ij}+\mathcal{T}_{ji}$. The graph is then a reflection of the back-and-forth trips performed by working individuals. While the graph produced is of directed type, in practice we can state that $w_{ij}=w_{ji}$ by construction.

\section{Acknowledgements}

AB acknowledges the financial support of the Ministerio de Economía y
Competitividad and the Ministerio de Ciencia e Innovación of Spain
under grant no. FJC2019-038958-I. VN acknowledges support from the
EPSRC New Investigator Award Grant No. EP/S027920/1.

\section{Data availability}

We obtained the Lazega lawyers friendship
network\cite{lazega2001collegial} from
\url{https://www.stats.ox.ac-.uk/~snijders/siena/Lazega_lawyers_data.htm}.
The social contact networks are available from the SocioPatterns
project\cite{genois2018can,sociopatterns}. We downloaded the income
and commuting data for London from
\url{https://data.gov.uk/dataset/d91901cb-13ee-4d00-906c-1ab36588943b/msoa-atlas}
and \url{http://www.nomisweb.co.uk/census/2011/wu03EW},
respectively. The Gowalla check-in data from
ref.\ \cite{liu2014exploiting} can be downloaded from
\url{http://www.yongliu.org/datasets/}. The power grid data set is
available for download from \url{https://github.com/mapequation/metadata-informed-community-detection-data}.

\appendix

\renewcommand{\thefigure}{S-\arabic{figure}}

\section{Additional results for contact networks}

We provide in Supplementary Figs.~\ref{InVS13} and~\ref{LH10}
additional results of the communities detected for a workplace
(InVS13) and a Hospital (LH10).

\begin{figure*}[!htbp]
  \begin{center}
    \includegraphics[width=6in]{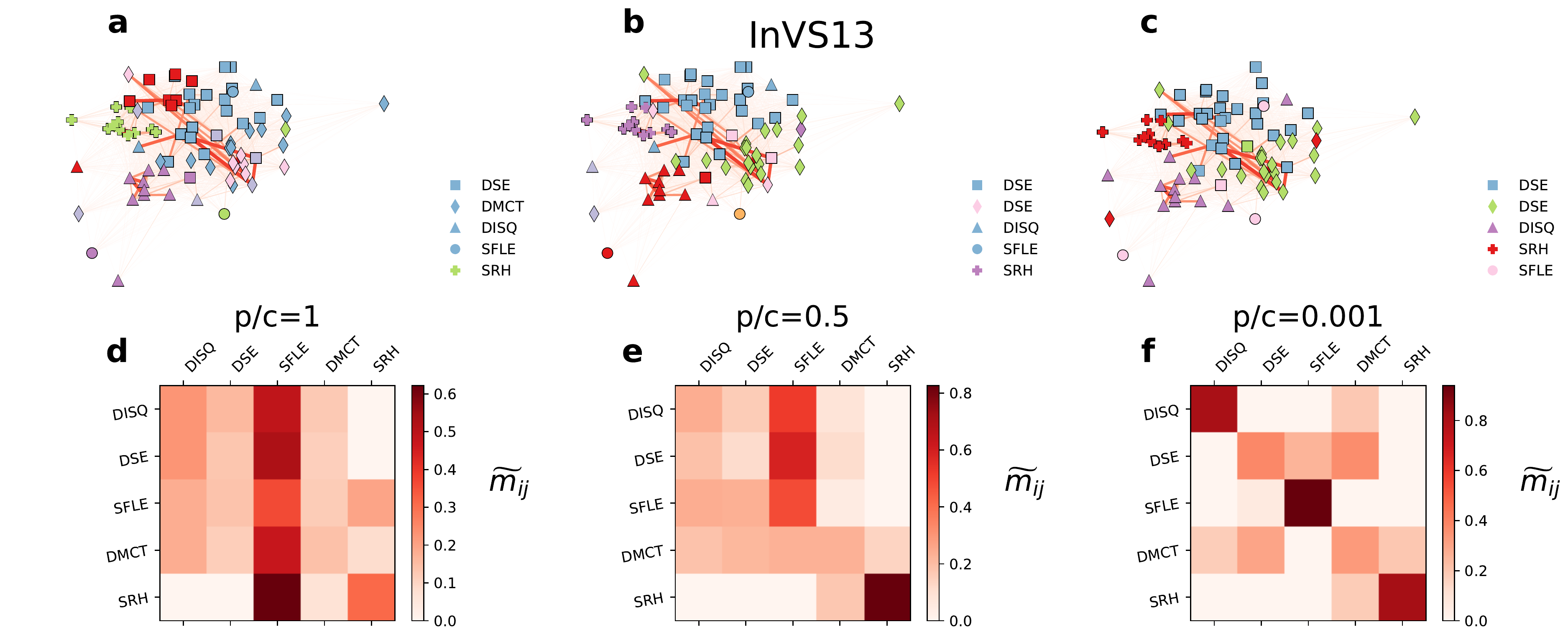}
  \end{center}
  \caption[Partitions obtained for the Workplace (InVS13) contact network.]{\textbf{Partitions obtained for the Workplace(InVS13) contact network.} Community detection analysis in the Lyon School contact graph where nodes correspond to individuals with a meta-data information. The probability to encode a transition is $p$ for nodes within the same class and $c$ otherwise. For a probability $p=1$, partitions when $c=1$ (\textbf{a}), $c=2$ (\textbf{b}) and $c=1000$ (\textbf{c}). \textbf{d-f} Class overlapping assignment $\widetilde{m}_{ij}$ when $c=1$ (\textbf{d}), $c=2$ (\textbf{e}) and $c=1000$ (\textbf{f}). Nodes are colored according to their community assignment while markers indicate their meta-data information.}\label{InVS13}
\end{figure*}

\begin{figure*}[!htbp]
  \begin{center}
    \includegraphics[width=6in]{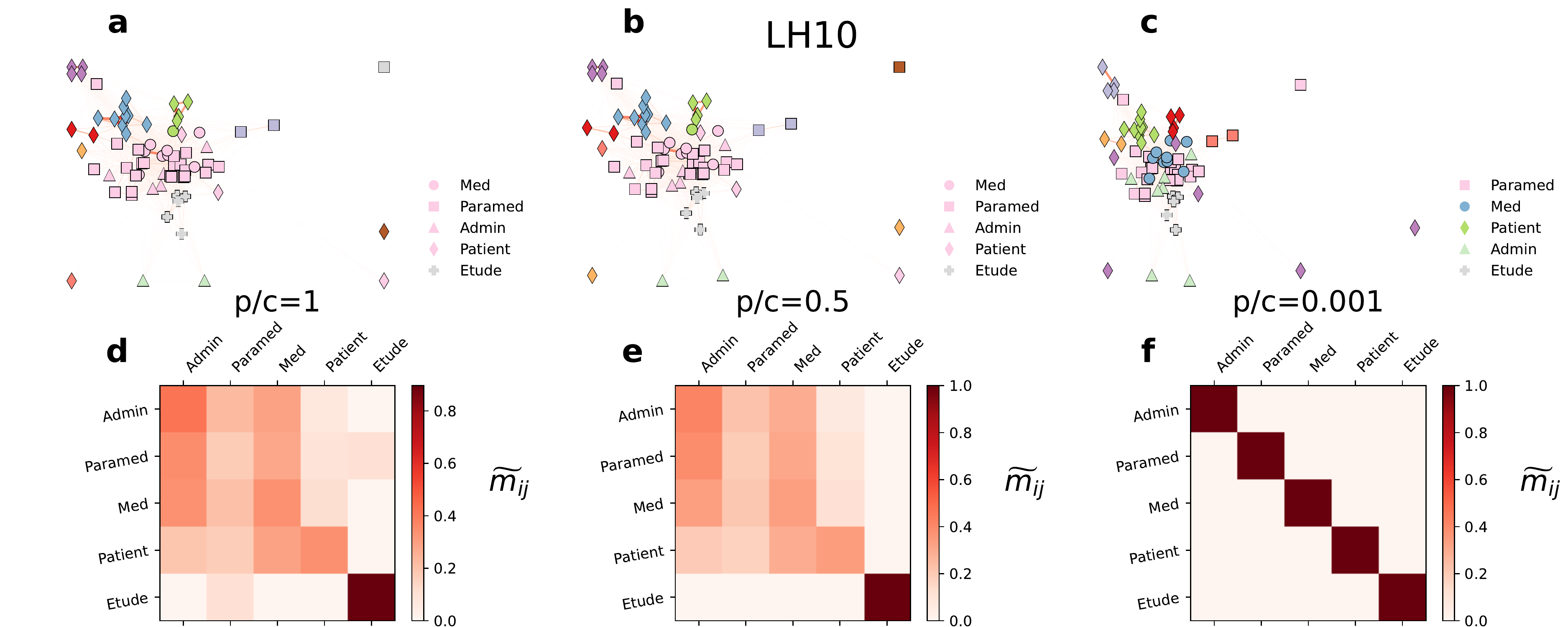}
  \end{center}
  \caption[Partitions obtained for the Hospital (LH10) contact network.]{\textbf{Partitions obtained for the Hospital (LH10) contact network.} Community detection analysis in the Hospital (LH10) contact graph where nodes correspond to individuals with a meta-data information. The probability to encode a transition is $p$ for nodes within the same class and $c$ otherwise. For a probability $p=1$, partitions when $c=1$ (\textbf{a}), $c=2$ (\textbf{b}) and $c=1000$ (\textbf{c}). \textbf{d-f} Class overlapping assignment $\widetilde{m}_{ij}$ when $c=1$ (\textbf{d}), $c=2$ (\textbf{e}) and $c=1000$ (\textbf{f}). Nodes are colored according to their community assignment while markers indicate their meta-data information.}\label{LH10}
\end{figure*}

\section{Additional results for urban activities}

\begin{figure*}[!htbp]
  \begin{center}
    \includegraphics[width=6in]{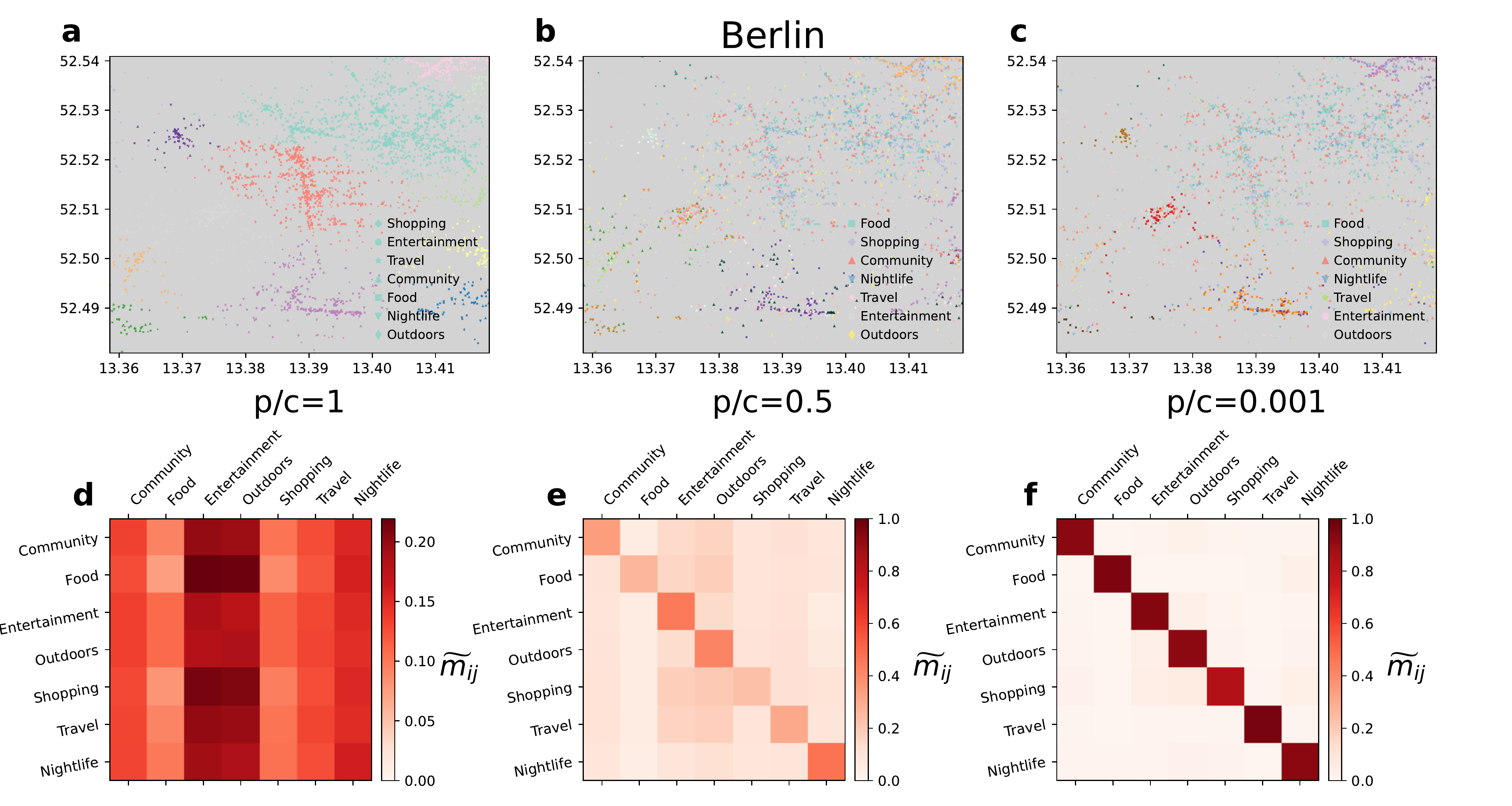}
  \end{center}
 \caption[Partitions obtained for the Gowalla venues in Prague]{\textbf{Partitions obtained for the Gowalla venues in Berlin.} Community detection analysis in the city of Berlin on the spatial graph connecting any pair of venues if their are closer than $2$ km. For a probability $p=0.5$, paritions when $c=1$ (\textbf{a}), $c=2$ (\textbf{b}) and $c=1000$ (\textbf{c}). \textbf{d-f} Class overlaping $\widetilde{m}_{ij}$ when $c=1$ (\textbf{d}), $c=2$ (\textbf{e}) and $c=1000$ (\textbf{f}). Venues are colored according to their community assignment and the marker indicated the type of venue.}\label{berlin}
\end{figure*}

\begin{figure*}[!htbp]
  \begin{center}
    \includegraphics[width=6in]{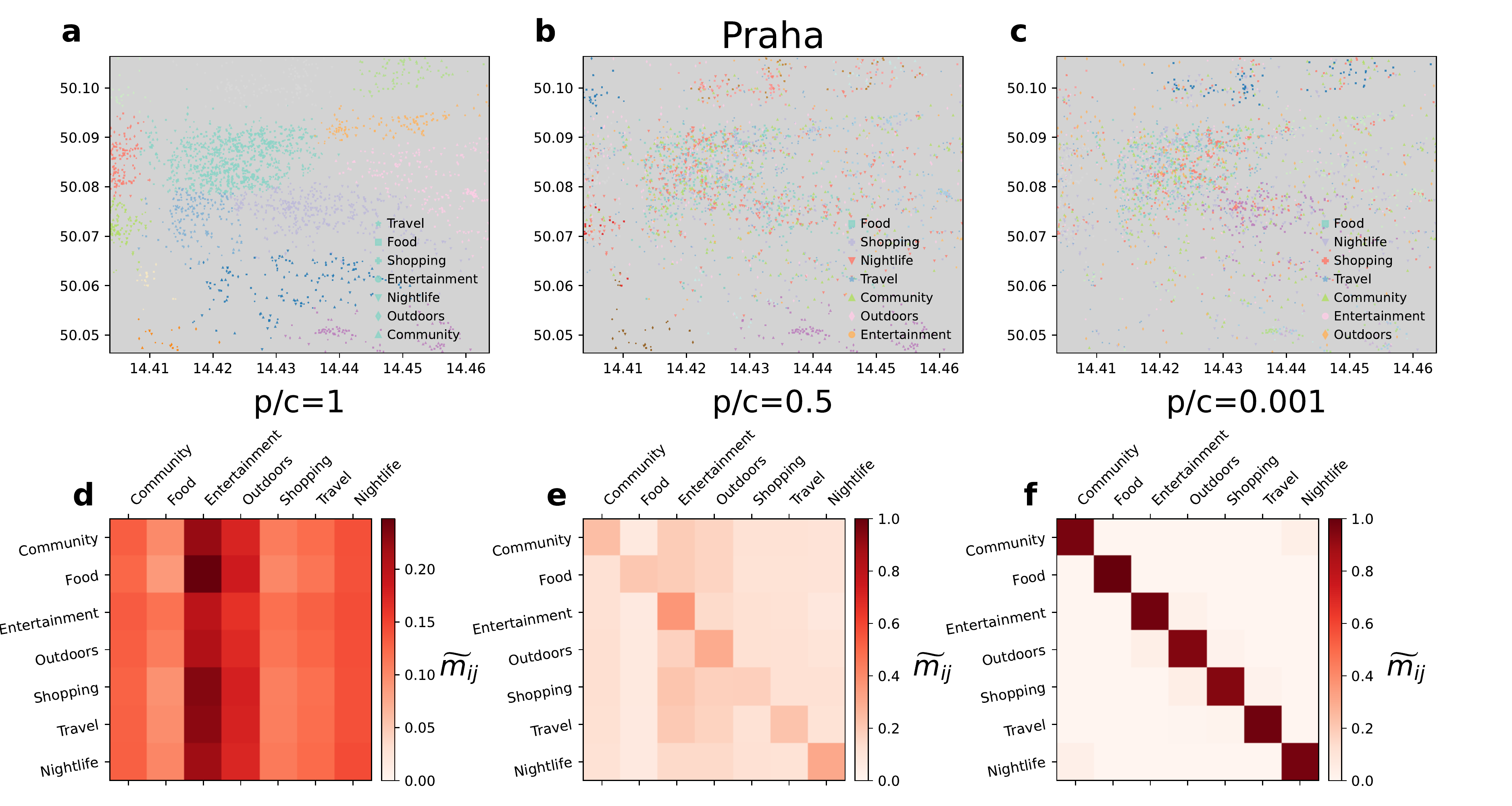}
  \end{center}
 \caption[Partitions obtained for the Gowalla venues in Prague.]{\textbf{Partitions obtained for the Gowalla venues in Prague.} Community detection analysis in the city of Prague on the spatial graph connecting any pair of venues if their are closer than $2$ km. For a probability $p=0.5$, paritions when $c=1$ (\textbf{a}), $c=2$ (\textbf{b}) and $c=1000$ (\textbf{c}). \textbf{d-f} Class overlaping $\widetilde{m}_{ij}$ when $c=1$ (\textbf{d}), $c=2$ (\textbf{e}) and $c=1000$ (\textbf{f}). Venues are colored according to their community assignment and the marker indicated the type of venue.}\label{prague}
\end{figure*}

In this section, we provide similar results on the spatial
clusterisation for urban activities in Berlin (Supplementary
Fig.~\ref{berlin}) and Prague (Supplementary Fig.~\ref{prague}).

\section{Additional results for the commuting network of London}

We provide in Supplementary Figs. \ref{unemployment}, \ref{lifeexpectancy}, \ref{deprivation} and \ref{obesity} additional results for the London commuting graph when classes are assigned according to unemployment, life expectancy, deprivation and obesity respectively.

\begin{figure*}[!htbp]
  \begin{center}
    \includegraphics[width=6in]{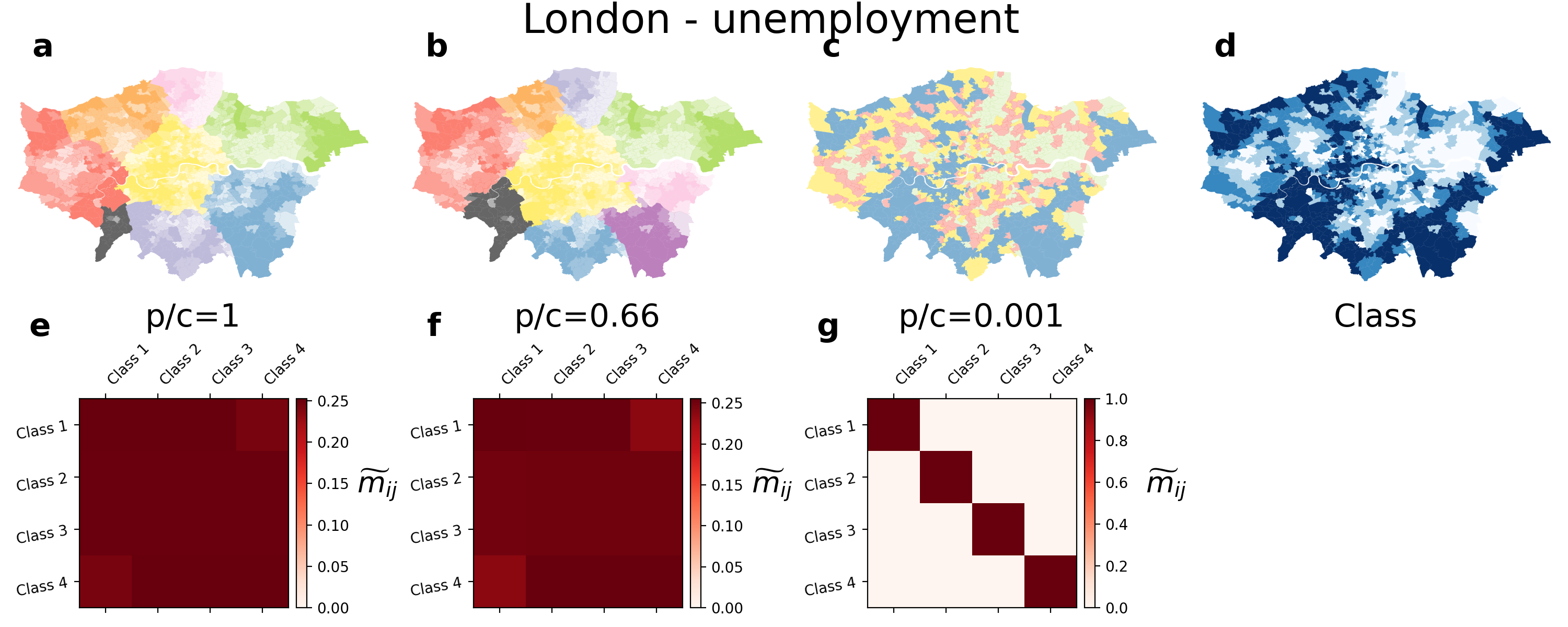}
  \end{center}
  \caption[Partitions obtained for the unemployment categories in the commuting network of London.]{\textbf{Partitions obtained for the unemployment categories in the commuting network of London.} Community detection analysis on the commuting network of London when the metadata is set according to the unemployment category. With regions in class 1 and 4 corresponding to the last and most wealthy, respectively. For a probability $p=1$, partitions when $c=1$ (\textbf{a}), $c=2$ (\textbf{b}) and $c=1000$ (\textbf{c}), with regions colored according to their community assignment. (\textbf{d}) Class assignment for each of the regions studied.  \textbf{e-g} Class overlapping  $\widetilde{m}_{ij}$ when $c=1$ (\textbf{e}), $c=1.5$ (\textbf{f}) and $c=1000$ (\textbf{g}). }\label{unemployment}
\end{figure*}

\begin{figure*}[!htbp]
  \begin{center}
    \includegraphics[width=6in]{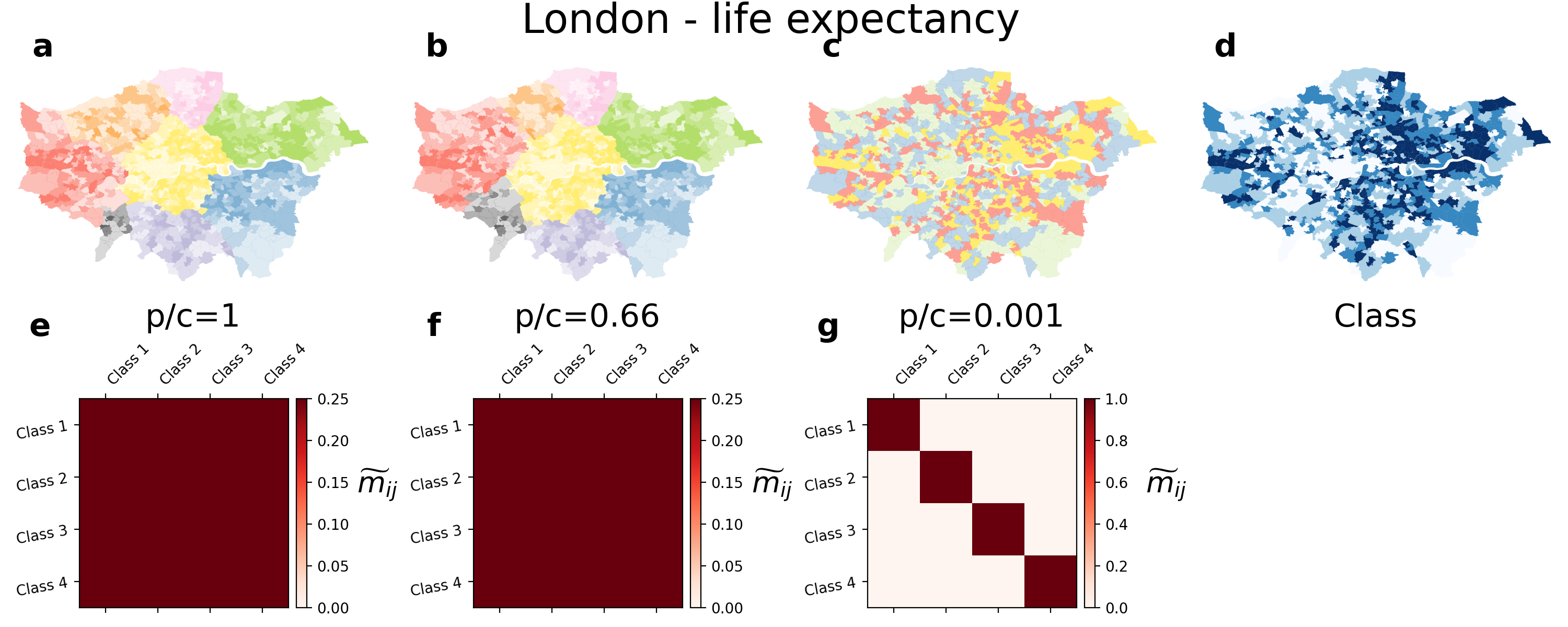}
  \end{center}
  \caption[Partitions obtained for the life expectancy categories in the commuting network of London]{\textbf{Partitions obtained for the life expectancy categories in the commuting network of London.} Community detection analysis on the commuting network of London when the metadata is set according to the life expectancy category. With regions in class 1 and 4 corresponding to the last and most wealthy, respectively. For a probability $p=1$, partitions when $c=1$ (\textbf{a}), $c=2$ (\textbf{b}) and $c=1000$ (\textbf{c}), with regions colored according to their community assignment. (\textbf{d}) Class assignment for each of the regions studied.  \textbf{e-g} Class overlapping  $\widetilde{m}_{ij}$ when $c=1$ (\textbf{e}), $c=1.5$ (\textbf{f}) and $c=1000$ (\textbf{g}). }\label{lifeexpectancy}
\end{figure*}

\begin{figure*}[!htbp]
  \begin{center}
    \includegraphics[width=6in]{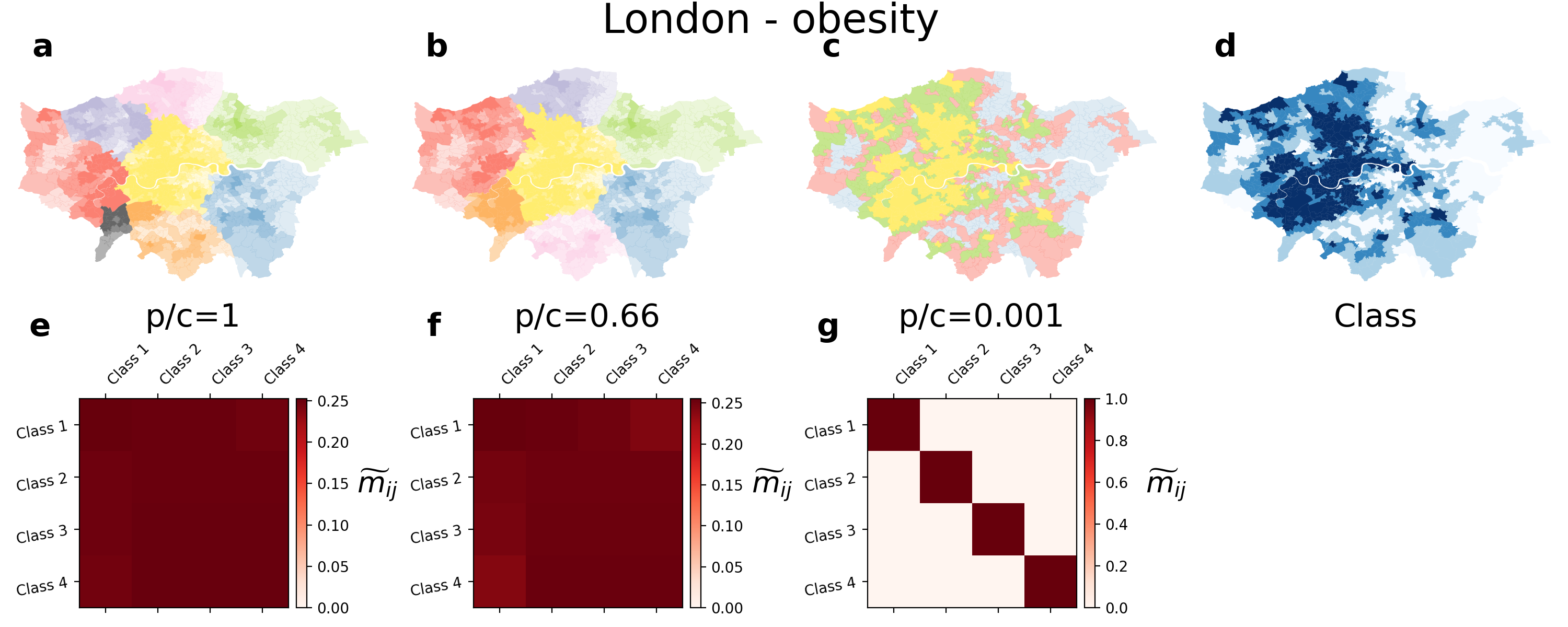}
  \end{center}
  \caption[Partitions obtained for the obesity categories in the commuting network of London]{\textbf{Partitions obtained for the obesity categories in the commuting network of London.} Community detection analysis on the commuting network of London when the metadata is set according to the obesity category. With regions in class Metadata-informed community detection with lazy encoding using absorbing random walks1 and 4 corresponding to the last and most wealthy, respectively. For a probability $p=1$, partitions when $c=1$ (\textbf{a}), $c=2$ (\textbf{b}) and $c=1000$ (\textbf{c}), with regions colored according to their community assignment. (\textbf{d}) Class assignment for each of the regions studied.  \textbf{e-g} Class overlapping  $\widetilde{m}_{ij}$ when $c=1$ (\textbf{e}), $c=1.5$ (\textbf{f}) and $c=1000$ (\textbf{g}). }\label{obesity}
\end{figure*}

\begin{figure*}[!htbp]
  \begin{center}
    \includegraphics[width=6in]{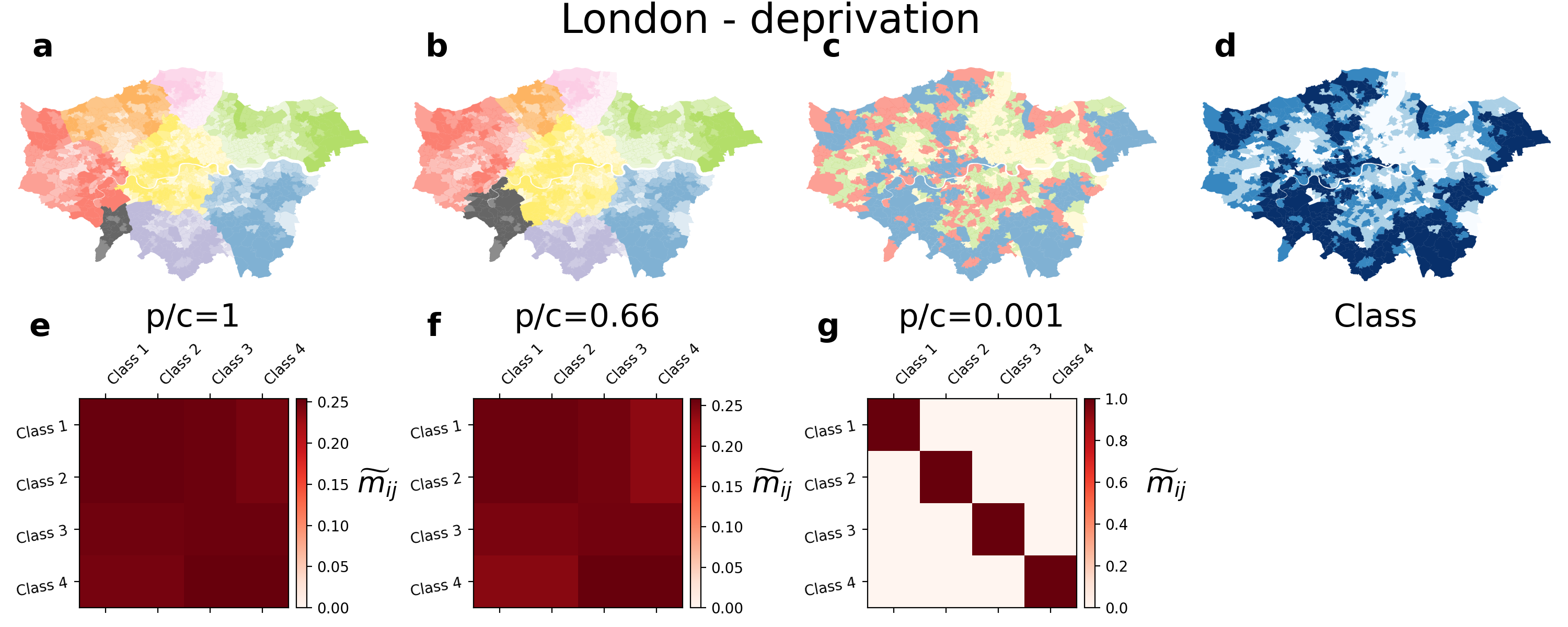}
  \end{center}
  \caption[Partitions obtained for the deprivation categories in the commuting network of London]{\textbf{Partitions obtained for the deprivation categories in the commuting network of London.} Community detection analysis on the commuting network of London when the metadata is set according to the deprivation category. With regions in class 1 and 4 corresponding to the last and most wealthy, respectively. For a probability $p=1$, partitions when $c=1$ (\textbf{a}), $c=2$ (\textbf{b}) and $c=1000$ (\textbf{c}), with regions colored according to their community assignment. (\textbf{d}) Class assignment for each of the regions studied.  \textbf{e-g} Class overlapping  $\widetilde{m}_{ij}$ when $c=1$ (\textbf{e}), $c=1.5$ (\textbf{f}) and $c=1000$ (\textbf{g}). }\label{deprivation}
\end{figure*}

\bibliography{references}

\end{document}